\documentclass[prd,preprint,superscriptaddress,showpacs]{revtex4}
\usepackage{revsymb}
\usepackage{graphicx}

\newcommand{\be}{\begin{equation}}
\newcommand{\ee}{\end{equation}}
\newcommand{\no}{\noindent}
\newcommand{\n}{\label}
\newcommand{\ke}{$k$-essence }
\newcommand{\kf}{$k$-field }

\newcommand{\ga}{\gamma_r}
\newcommand{\T}{$T_r$-field }
\newcommand{\Ts}{$T_r$-fields }

\begin{document}

\title{Extended tachyon field, Chaplygin gas and solvable \ke cosmologies}

\author{Luis P. Chimento}
\email{chimento@df.uba.ar}
\affiliation{Departamento de F\'{\i}sica,
Facultad de Ciencias Exactas y Naturales,
Universidad de Buenos Aires,
Ciudad  Universitaria,  Pabell\'on  I,
1428 Buenos Aires, Argentina.}[]

\bibliographystyle{plain}

\begin{abstract}

We investigate a flat Friedmann-Robertson-Walker spacetime filled with \ke and
find the set of functions $F$ which generate three different families of
extended tachyon fields and Chaplygin gases. They lead to accelerated and
superaccelerated expanding scenarios.

For any function $F$, we find the first integral of the \kf equation when the
\kf is driven by an inverse square potential or by a constant one. In the
former, we obtain the general solution of the coupled Einstein-\kf equations
for a linear function $F$. This model shares the same kinematics of the
background geometry with the ordinary scalar field one driven by an
exponential potential. However, they are dynamically different. For a constant
potential, we introduce a \kf model that exhibits a transition from a
power-law phase to a de Sitter stage, inducing a modified Chaplygin gas.

\end{abstract}

\vskip 1cm


\maketitle

\newpage

\date{\today}

%==================================================================
\section{Introduction}
%==================================================================

Cosmological inflation has become an integral part of the standard model of
the universe and perhaps the only known mechanism which can dynamically solve
the flatness and the horizon problem of the universe \cite{li}. It has gained
certain support from the recent observations of the cosmic microwave
background anisotropies \cite{be}-\cite{ha}. Although particle physics, in
particular M/String theory, provides several very weakly coupled scalar fields
which are natural inflaton candidates, there exists no clearly preferred
inflationary model. A link between string theory and inflation was
investigated in Ref. \cite{ar3} where the authors introduce the $k$-inflation
and show that the $k$-field may drive an inflationary evolution starting from
rather generic initial conditions. Also, the process of rolling tachyon field
has been extensively used to implement power-law accelerated expansion
scenarios \cite{a}-\cite{b}. In Ref. \cite{sen1} it was pointed out that
rolling tachyons can contribute a mass density to the universe that resembles
classical dust. This has brought a new understanding of the role of the
tachyon in string theory. The tachyonic matter, then, might provide an
explanation for inflation at the early epochs and could contribute to some new
form of cosmological dark energy \cite{acce} at late times \cite{gib}. These
facts suggest to enlarge the theoretical background exploring some new
possibilities, in this sense, we will introduce an extended tachyon field
(ETF) which supply accelerated and superaccelerated expanding scenarios.

In Ref. \cite{brief} it was argued that the coincidence problem may be solved
assuming a universe filled with a viscous fluid and dark energy modelled with
a tachyonic field or a Chaplygin gas. The conjecture that dark matter and dark
energy can be unified by using a generalized Chaplygin gas obeying an exotic
equation of state has been investigated in several works in view of the
cosmological observations \cite{bento-03}-\cite{cat}. The present situation is
somewhat controversial, with some tests indicating good agreement between
observational data and the theoretical predictions of the model and others
ruling out the model as an actual possibility of description for our Universe
\cite{zal}. There were found some differences with the observed CMB and mass
power spectrum data . Due to these discrepancies, it may be useful to consider
other candidates for dark energy, as for instance, the extended Chaplygin gas
(ECG) generated from the ETF. Another interesting possibility appears
selecting a $k$-field model whose equation of state is like that of two
fluids, one obeying a baryotropic equation of state with constant baryotropic
index $\gamma$, and the other is an ECG. This "modified Chaplygin gas"
interpolates between a power-law phase and a de Sitter phase.

From theoretical and experimental points of view it is important to find an
exact shape of the potential, for instance, tachyonic inflation has been
studied using phenomenological potentials that have not been derived from
string theory and can be related to the so called ``k-Inflation'' \cite{ar3},
\cite{alex,padma}. Such $k$-fields, described by a non-standard kinetic term,
has been one of the recent suggested candidates to play the role of same
unknown component of the universe. One of the purpose of introducing \ke is to
provide a dynamical explanation of cosmological observations which does not
require the fine-tuning of initial conditions \cite{ar2}-\cite{mal1}. In this
sense, it may be considered as an alternative to quintessence, which requires
a scalar field that slowly rolls down a potential to depict the observed
acceleration of the present universe. Also, it was argued that in certain
dynamical regimes the $k$-essence is equivalent to quintessence and it may
prove difficult to distinguish between the two fields. In the light of these
results, it seems clear that it is worth to search the links between scalar
field and \ke cosmologies, in especial, whether they are kinematically or
dynamically equivalent.

In section II, we introduce the ETF and find the power-law solution generated
by an inverse square potential. At the end, we define the ECG and show the
general behaviour of the background geometry. In section III, we find the
first integral of the \kf equation for an inverse square potential and the
general solution of coupled Einstein-\kf equations for a linear function $F$.
For power-law solutions we demonstrate that the linear \kf model, with
constant function $F$, is isomorphic to the model with divergent sound speed.
Finally, we introduce the modified Chaplygin gas. In section IV, we show the
kinematical equivalence of the $k$-field driven by a tachyonic potential and a
linear function $F$, with the scalar field driven by an exponential potential.
The conclusions are stated in section V.

%==================================================================
\section{Extend tachyon field and Chaplygin gas}
%==================================================================

The spatially flat homogeneous and isotropic space-time is described by the
Friedmann-Robertson-Walker (FRW) line element

\begin{equation}
\label{metric}
ds^{2} =-dt^{2}+
a^{2}(t)\left[dx^{2}_{1}+dx^{2}_{2}+dx^{2}_{3}\right],
\end{equation}

\noindent where $a(t)$ is the scale factor. This metric allows a particular
calculational simplicity, on account of both the high degree of symmetry and
the single metric degree of freedom. We assume that $8\pi G=1$.

Let us present the Lagrangian of the \ke field

\be
\n{L}
{\cal L}=-V(\phi)\,F(x),   \qquad   x=g^{ik}\phi_i\phi_k,
\ee

\no where $V(\phi)$ is a positive defined potential, $F(x)$ is an arbitrary
function of $x$, $\phi$ is the $k$-field and $\phi_i=\partial V(\phi)/\partial
x^i$. Associating the energy-momentum tensor of the $k$-field,

\be
\n{t}
T_{ik}=V(\phi)\left[2F_x\phi_i\phi_k-g_{ik}F\right],
\qquad F_x=\frac{d\,F}{d\,x},
\ee

\no with the energy-momentum tensor of a perfect fluid, we compute the energy
density $\rho$, the pressure $p$ and the baryotropic index $\gamma=1+p/\rho$
of this equivalent fluid

\be
\n{ro}
\rho=V(\phi)[F-2xF_x], \qquad  p={\cal L}=-V(\phi)F,
\ee

\be
\n{ga}
\gamma=-\frac{2\dot H}{3H^2}=-\frac{2xF_x}{F-2xF_x}.
\ee

\no From Eq. (\ref{ro}), the Einstein field equations are

\be
\n{00}
3H^2=V[F-2xF_x],
\ee

\be
\n{11}
\dot H=VxF_x,
\ee

\no and the conservation equation reads

\be
\n{con}
\dot\rho+3H(\rho+p)=0,
\ee

\no where $H=\dot a/a$ is the expansion rate. Substituting the Eq. (\ref{ro})
into the conservation equation (\ref{con}), we find the field equation for
the \kf

\be
\n{kg}
[F_x+2xF_{xx}]\ddot\phi+3HF_x\dot\phi+\frac{V'}{2V}[F-2xF_x]=0.
\ee

The stability of the \ke with respect to small wavelength perturbations
requires that the effective sound speed

\be
\n{c}
c_s^2=\frac{p_x}{\rho_x}=\frac{F_x}{F_x+2xF_{xx}},
\ee

\no be positive. However, in Ref. \cite{car} it was shown that a positive
sound speed is not a sufficient condition for the theory to be stable. For the
tachyon field, which is obtained from the \kf selecting $F=(1-\dot
T^2)^{1/2}$, with $\dot T^2=-x=\dot\phi^2$ the sound speed is
$c_s^2=1-\gamma>0$, where $\gamma=\dot T^2$. Let us look for the set of
functions $F$, such that, $c_s^2$ is proportional to the sound speed of the
tachyon field. We express this proportionality as

\be
\n{=}
c_{s}^2=\frac{1-\gamma}{2r-1},
\ee

\no where $r$ is a real constant, with $r<1/2$ for $\gamma>1$ and $r>1/2$ for
$\gamma<1$. From Eqs. (\ref{ga}), (\ref{c}) and (\ref{=}), we obtain a
differential equation for the function $F$

\be
\n{==}
(1-r)FF_x+(2r-1)xF_x^2+xFF_{xx}=0.
\ee

\no Integrating for $r=0$, we obtain the set of functions
$F_0(x)=x^{\gamma_0/2(\gamma_0-1)}$, with $\gamma_0$ constant, investigated in
\cite{al}; while for $r=1$ we find three types of solution $F_1=(1-\dot
T_1^2)^{1/2}$, $F_1=(1+\dot T_1^2)^{1/2}$, and $F_1=-(\dot T_1^2-1)^{1/2}$.
The first function corresponds to the ordinary tachyon field $T=T_1$. Its
energy density $\rho_1=VF_1^{-1}$ and pressure $p_1=-VF_1$ leads to the
relation $p_{1}=(\gamma_1-1)\rho_1$, where $\gamma_1=\dot T_1^2$. The other
two functions generate two "new tachyon field" which will be investigated
below. For the remaining values of $r$ the general solution of Eq. (\ref{==})
is given by $F_r^{2r}=c_1(-x)^r+c_2$, where $c_1$ and $c_2$ are arbitrary
integration constants, the baryotropic index (\ref{ga}) is
$\gamma_r=-c_1(-x)^r/c_2$ and we have $F_r^{2r}=c_2(1-\ga)$. From Eqs.
(\ref{ro}) and (\ref{ga}), the energy density of the \kf can be rewritten as
$\rho=VF/(1-\gamma)$. So, without loss of generality, we can split the
solutions as follow: for $\ga<1$ and $r>1/2$, it is necessary that $F_r>0$
and $c_2>0$, with two options, $0<\ga<1$ and $c_1<0$ or $\ga<0$ and $c_1>0$;
for $\ga>1$ and $r<1/2$, we have $F_r<0$, $c_2$ and $c_1>0$. Finally

\be
\n{fo}
F_r=(1-\ga)^{1/2r},   \qquad   \ga=\pm\,\dot T_r^{2r}<1,
\ee

\be
\n{fe}
F_r=-(\ga-1)^{1/2r},   \qquad   \ga=\,\dot T_r^{2r} >1,
\ee

\no where $\ga$ is the extended baryotropic index and the \T linked with $F_r$
will be called ETF. Inserting Eqs. (\ref{fo}) and (\ref{fe}) into (\ref{ro}),
we get the energy density $\rho_r$ and the pressure $p_r$ of each ETF

\be
\n{ro1}
\rho_r=V(1-\dot T_r^{2r})^{(1-2r)/2r}, \qquad
p_r=-V(1-\dot T_r^{2r})^{1/2r}, \qquad 0<\ga<1,
\ee

\be
\n{ro2}
\rho_r=V(1+\dot T_r^{2r})^{(1-2r)/2r}, \qquad
p_r=-V(1+\dot T_r^{2r})^{1/2r}, \qquad  \ga<0,
\ee

\be
\n{ro3}
\rho_r=V(\dot T_r^{2r}-1)^{(1-2r)/2r}, \qquad
p_r=V(\dot T_r^{2r}-1)^{1/2r}, \qquad  1<\ga,
\ee

\no with

\be
\n{pn}
p_r=(\ga-1)\rho_r.
\ee

\no The Eq. (\ref{ro1}) comprise perfect fluids which generalize the normal
tachyon field, which is obtained for $r=1$. The fluids represented by the Eq.
(\ref{ro2}) with negative pressure and negative baryotropic index describe
phantom cosmologies, while those described by the Eq. (\ref{ro3}) give rise to
nonaccelerated expanding evolutions. In the limit of large $r$, the fluids
(\ref{ro1}) and (\ref{ro2}) satisfy the equation of state $p=-\rho=-V$ acting
like a variable cosmological constant depending on the k-field. Besides, the
fluids (\ref{ro3}) fulfill the equation of state $p=\rho=V$ behaving like
stiff matter in the same limit. These exotic fluids satisfy the relations

\be
\n{ch-}
p_r=-\frac{V^{2r/(2r-1)}}{\rho_r^{1/(2r-1)}},  \qquad  \ga<1,
\ee

\be
\n{ch+}
p_r=\frac{V^{2r/(2r-1)}}{\rho_r^{1/(2r-1)}},   \qquad  \ga>1,
\ee

\no which, for a constant potential $V=V_0$, become exotic equations of state
which extend that of generalized Chaplygin gas.

%--------------------------------------------------------------------------
\subsection{Power-law expansion for the \T}
%--------------------------------------------------------------------------

We begin investigating the atypical behaviour of the $T_{1/2}$-field. Its
sound speed diverges and the \kf equation (\ref{kg}) becomes a first order
equation. The equation (\ref{ga}) can be integrated and its solution gives a
link between the expansion rate and $T_{1/2}$

\be
\n{mi}
H=\frac{2}{3T_{1/2}},
\ee

\no where the integration constant has been chosen to set the singularity at
$T_{1/2}=0$. For $r=1/2$, the three equations (\ref{ro1}), (\ref{ro2}) and
(\ref{ro3}) lead to $3H^2=V$. So, combining them with Eq. (\ref{mi}) we obtain
the inverse square potential

\be
\n{pm}
V_{1/2}=\frac{4}{3T_{1/2}^2},
\ee

\no which is the uniquely allowed potential for $T_{1/2}$. As $T_{1/2}$ is not
controlled by the \kf equation (\ref{kg}), it can be chosen freely and this
choice determines, after integrating Eq. (\ref{mi}), the form of the scale
factor. Although the potential (\ref{pm}) diverges at $T_{1/2}=0$, it
reasonably mimics the behaviour of a typical potential in the condensate of
bosonic string theory. One expects the potential to have a unique local
maximum at the origin and a unique global minimum away from the origin at
which $V$ vanishes. In the most interesting case the global minimum is taken
to lie at infinity. Obviously more complicated potentials may be contemplated
but this is the simplest case to begin with.

It will be useful to investigate the existence of accelerated and
superaccelerated expanding solutions when the \T is driven by the potential
$V_r=V_0/T_r^2$. To do that, let us consider an evolution of the form $a=t^n$
and a $T_r$-field, such that $T_r\propto t$. Power law solutions are very
important because they can be always obtained from any function $F$ with an
inverse square potential or from polynomial functions $F=(-x)^N$ with any
potential. As these solutions are usual ingredients in the quintessense and \ke
models they would allow to recognize and compare the differences between the
two cosmological models. The complete solution for the ETF is given by

\be
\n{st1}
T_r=\left(\frac{2}{3n}\right)^{1/2r}t,
\qquad n=\frac{1}{3}\left[1+\sqrt{1+9\beta^2}\right],
\qquad 0<\ga<1,
\ee

\be
\n{st2}
T_r=\left(\frac{2}{-3n}\right)^{1/2r}t,
\qquad  n=\frac{1}{3}\left[1-\sqrt{1+9\beta^2}\right],
\qquad \ga<0,
\ee

\be
\n{st3}
T_r=\left(\frac{2}{3n}\right)^{1/2r}t,
\qquad  n=\frac{1}{3}\left[1\pm\sqrt{1-9\beta^2}\right],
\qquad 1<\ga,
\ee

\no where

\be
\n{b}
\beta^2=\left[\frac{3}{V_0}\left(\frac{2}{3}\right)^{1/r}\right]^{2r/(1-2r)}.
\ee

\no The sound speed associated with the above solutions is given by

\be
\n{ct}
c_s^2=\frac{1-2/3n}{2r-1}.
\ee

\no Assuming a positive sound speed bounded by the light speed $0\le c_s^2\le
1$ we can see that $r>1-1/3n$ is required for $n>2/3$ or $n<0$ and $r<1-1/3n$
for $0<n<2/3$.

An accelerated power-law expanding universe ($n>1$), can be described by the
set of ETF having $r>2/3$ and it is represented by the solutions (\ref{st1}).
All these models are cinematically and dynamically different because the scale
factor and the \kf are strongly depending of $r$ as can be seen from Eqs.
(\ref{st1}) and (\ref{b}). In particular, for $r=1$ the Eq. (\ref{st1})
reduces to the solution of the tachyon field found in \cite{alex,padma}. The
solutions (\ref{st2}) with $n<0$ describe phantom cosmologies and we have to
select ETF with $r>1$ to reach this scenario
 
%--------------------------------------------------------------------------
\subsection{Extended Chaplygin gas}
%--------------------------------------------------------------------------

Recently it was proposed a set of simple cosmological models based on the use
of peculiar perfect fluids \cite{kam}. In this simple model the universe is
filled with the so called Chaplygin gas, which is a perfect fluid
characterized by the following equation of state $p = -A/\rho$, where $A$ is a
positive constant. It describes a transition from a decelerated cosmological
expansion to a cosmic accelerated de Sitter stage. Other possibility is the
inhomogeneous Chaplygin gas, which is able to combine the roles of dark energy
and dark matter \cite{bi-02}, and the generalized Chaplygin gas model
discussed in Ref. \cite{bento-02}, having two free parameters
$p=-A/\rho^{\alpha}$ with $0<\alpha\leq 1$.

These cosmological models may give a unified macroscopic phenomenological
description of dark energy and dark matter and generalize the usual
$\Lambda$CDM models. On the other hand, the Chaplygin gas can be considered as
the simplest tachyon cosmological models where the tachyon field is a purely
kinetic \ke model with a constant potential. In the same way, we will show
that the generalized Chaplygin gas can be conceived as the simplest ETF model
driven by a constant potential. This identification has the advantage of
producing a variety of new Chaplygin gases, same of which leads to
superaccelerated scenarios as we will see in this subsection.

Coming back to our Eqs. (\ref{ch-}) and (\ref{ch+}), we see that in the
case of a constant potential $V(T_r)=V_0$, with $r=1$ and $0<\ga<1$, the
exotic equation of state (\ref{ch-}) represents a Chaplygin gas \cite{kam}.
Such equation may be the consequence of a scalar field with a non-standard
kinetic term, e.g., the string theory motivated tachyon field \cite{sen,gibb}.
For $1<r$ and $0<\ga<1$, the equation of state (\ref{ch-}) represents a perfect
fluid called generalized Chaplygin gas \cite{bento-02}. Finally an ECG will be
characterized by the equation of state (\ref{ch-}) with $1/2<r<1$, or by the
Eq. (\ref{ch+}) with $r<1/2$. Their properties will be investigated below.
Using Eqs. (\ref{ch-})-(\ref{ch+}) and the relativistic energy conservation
equation (\ref{con}), we obtain the energy density of the ECG:

\be
\n{ro1c}
\rho=V_0\left[1+\left(\frac{a_0^3}{a^3}\right)^{2r/(2r-1)}\right]^{(2r-1)/2r},
\qquad 0<\ga<1,
\ee

\be
\n{ro2c}
\rho=V_0\left[1-\left(\frac{a_0^3}{a^3}\right)^{2r/(2r-1)}\right]^{(2r-1)/2r},
\qquad \ga<0,
\ee

\be
\n{ro3c}
\rho=V_0\left[-1+\left(\frac{a_0^3}{a^3}\right)^{2r/(2r-1)}\right]^{(2r-1)/2r},
\qquad 1<\ga.
\ee

\no The Eq. (\ref{ro1c}) with $r>1$ gives the energy density of the
generalized Chaplygin gas in terms of the scale factor interpolating between a
dust dominated phase where the energy density is $\rho\approx V_0(a_0/a)^3$,
and a de Sitter phase where $\rho\approx V_0$. While during the intermediate
stage it could be interpreted as a mixture of two-fluids, one of which is the
cosmological constant and the other is a perfect fluid with equation of state
$p\propto\rho$. The additional free parameter $r$ of the generalized Chaplygin
gas can be used to compare it with observational data.

In any other case, Eqs. (\ref{ro1c}) with $1/2<r<1$, (\ref{ro2c}) and
(\ref{ro3c}) represent new perfect fluids which are interesting from the
cosmological point of view. The scale factor generated by the source
(\ref{ro2c}) is nonsingular and has a bounce at the minimum $a=a_0$. The
universe begins from a contracting era and ends in a superaccelerated stage.
Such cosmologies may be interpreted as  universes filled with baryotropic
fluids having a negative constant baryotropic index and violating the weak
energy condition $\rho>0$, $\rho+p>0$. The models will  be dubbed phantom
cosmologies following the standard terminology. Phantom matter can apparently
be accommodated by current observations \cite{obs}, and it can be based on the
motivation provided by string theory \cite{str}. It looks interesting to admit
that the origin of dark energy should be searched within a fundamental theory,
as string theory. However, at present there is no consensus whether a universe
violating the weak energy condition should generically possess a future
singularity or big rip \cite{fate}.

The source (\ref{ro3c}) leads to two types of solutions according to the
values of the parameter $r$. For $r<0$ the universe has a finite time span,
interpolating between two dust dominated phases and has a maximum at $a=a_0$.
For $0\le r\le 1/2$ the universe begins at a singularity with a finite scale
factor, $a=a_0$, and ends in a dust dominated phase. From the cosmological
point of view these solutions are not relevant because they do not describe
the present observed accelerated expansion stage.

%==============================================================
\section{Solvable k-essence cosmologies}
%==============================================================

In this section, we will show some cases where the coupled Einstein-\kf
equations (\ref{00},\ref{kg}) admit a first integral or can be solved exactly.
Expressing the energy density of the \kf as $\rho=VF/(1-\gamma)$ and using the
conservation equation (\ref{con}), we get the \kf equation in terms of the
baryotropic index $\gamma$

\be
\n{ga.}
\left(\frac{\gamma}{\dot\phi}\right)^.+
3H\left(\frac{\gamma}{\dot\phi}\right)(1-\gamma)+\frac{V'}{V}(1-\gamma)=0.
\ee

\no where $V'=dV/d\phi$. This form of writing the field equation allows to
show that the first integral of the \kf equation (\ref{kg}) for any function
$F$ is giving by

\be
\n{pig}
\frac{\gamma}{\dot\phi}=\phi^{-1}\left(\frac{2}{3H}+\frac{c}{a^3H^2}\right),
\ee

\no where we have ssumed an inverse square potential

\be
\n{V}
V=\frac{V_0}{\phi^2},
\ee

\no and $c$ is an arbitrary integration constant. The usual linear field
solution $\phi=\phi_0 t$ along with the power law scale factor
$a=t^{2/3\gamma}$ (with constant $\gamma$) is obtained from the last equation
for a vanishing integration constant. Combining Eqs. (\ref{ga}), (\ref{11})
and (\ref{V}) with Eq. (\ref{pig}), it can be rewritten as

\be
\n{kgp}
\dot\phi F_x-\left[H+\frac{3c}{2a^3}\right]\frac{\phi}{V_0}=0,
\ee

\no or as

\be
\n{hp}
-V_0F_x\dot H=\left[H+\frac{3c}{2a^3}\right]^2.
\ee

\no Then, for the tachyonic potential (\ref{V}), Eqs. (\ref{pig}),
(\ref{kgp}), (\ref{hp}) are three different forms of writing the first
integral of the \kf equation (\ref{kg}) or (\ref{ga.}).

At this point, using Eqs. (\ref{ga}), (\ref{V}) and (\ref{pig}), it is
interesting to see the coupled Einstein-\kf equations (\ref{00},\ref{kg}) as a
system of differential equations for the function $F$. Hence, integrating we
obtain

\be
\n{i2}
F=\frac{3h_0^2}{V_0}+\sqrt{-x}\left[b+\frac{3c}{2V_0}
\int\frac{(2h_0+ch)h}{(-x)^{3/2}}dx\right],
\ee

\be
\n{i1}
H\phi=-\frac{3c}{2}\int\frac{d\phi}{a^3}=h_0+ch,
\ee

\no where $h_0$ and $b$ are arbitrary constants and $\dot h=-3\dot\phi/2a^3$.
In the particular case $c=0$, the Eq. (\ref{i2}) becomes

\be
\n{finf}
F^\infty=\frac{3h_0^2}{V_0}+b\sqrt{-x},
\ee
 
\no which, after a redefinition of the constants, turns into the function
$F_{1/2}$ that generates the ETF, $T_{1/2}$, (see Eqs. (\ref{fo},\ref{fe}) for
$r=1/2$). For this "divergent" \ke theory the sound speed (\ref{c}) diverges
and the \kf equation (\ref{kg}) becomes a first order equation that is only
consistent with an inverse square potential. This divergent model is related
to the linear \kf model, $\phi=\phi_0\,t$, obtained by evaluating $F$, $F_x$
at $x=x_0=-\phi_0^2$ into the Einstein-$k$-field equations
(\ref{00},\ref{kg}). The linear \kf model driven by the potential (\ref{V})
leads to the power-law solutions $a=t^\lambda$ with

\be
\n{sol}
\lambda=\frac{1}{3}\frac{f+2\phi^2_0f'}{\phi^2_0f'}, \qquad
V_0=\frac{\lambda}{f'},
\ee
 
\no where we have defined $f=F(-\phi^2_0)$, $f'=F_x(-\phi^2_0)$.

From Eqs. (\ref{finf},\ref{sol}) it is easy to show that the linear model is
isomorphic to the divergent one. This can be done by constructing a one-to-one
mapping between these two models. In fact, choosing in the divergent model the
constants $h_0^2=\lambda^2\phi^2_0$ and $b=-2\lambda\phi_0/V_0$ we find the
same power-law solutions obtained from the linear model. In Ref. \cite{al} it
was suggested that this might be the reason as to why a model with a diverging
sound speed leads to serious problems as discussed in a recent paper
\cite{mal2}. In addition, assuming a series expansion of the function $F(x)$
around $x=x_0$, the background cosmology is completely determined by the first
two coefficients $(f, f')$ of the expansion of $F$ and the value of $\phi_0$.
Hence, the model is insensitive to the remaining coefficients in the expansion
of the function $F$ and both the linear and divergent models should be
considered as equivalent. This means that the power-law solutions generated by
an inverse square potential possesses a degeneracy. Possibly this degeneracy
may be removed by perturbating the solutions.

%==================================================================
\subsection{Inverse square potential and linear function $F$}
%==================================================================

From the \kf equation (\ref{hp}), it can be seen that a linear function

\be
\n{fg}
F=1+mx,
\ee

\no where $m$ is a constant, decouples the dynamics of the background geometry
from the dynamics of the \kf. Therefore, this choice clearly introduces a
rather small degree of nonlinearity into the dynamical equations allowing us
to obtain the general solution of the coupled Einstein-\kf equations. On the
other hand, the function (\ref{fg}) mimics the behaviour of other models. For
instance, when $x\ll 1$ the tachyonic function $F=(1+x)^{1/2}$ can be
approximated by $F\approx 1+x/2$ and it has the form (\ref{fg}) \cite{padma}.
In Ref. \cite{ar1} it was introduced a set of models where $F$ admits a power
series expansion similar to (\ref{fg}). This form is reminiscent of a
Born-Infeld action with higher order corrections in $x$, and particular cases
were investigated in \cite{mal1,mal2}. Hence, the knowledge of the general
solution for the linear function (\ref{fg}) should be of interest, at least
for understanding the asymptotic behaviour of many others models generated by
an analytical function $F(x)=F(0)+F_x(0)x+......$. In fact, keeping the first
order term in the expansion of $F$, it adopts the linear form (\ref{fg}),
after redefining the potential $V$ to set the constant $F(0)=1$.

Combining Eqs. (\ref{hp}) and (\ref{fg}), we obtain the following nonlinear
second-order differential equation for the scale factor

\begin{equation}
\label{s..}
\frac{d^2s}{d\tau^2}+s^{\sigma}
\frac{ds}{d\tau}+\frac{1}{4}s^{2\sigma+1}=0, \qquad  \sigma=-3mV_0,
\end{equation}

\noindent where we have used the new variables $s$ and $\tau$, defined by

\begin{equation}
\label{stau}
s=a^{-3/\sigma},   \qquad      \tau=\frac{3c}{mV_0}t.
\end{equation}

\no The general solution of Eq. (\ref{s..}) can be found changing to the
nonlocal variables $z$ and $\eta$, defined by \cite{lmath}

\begin{equation}
\label{zeta}
z=\frac{s^{\sigma+1}}{\sigma+1},   \qquad   \eta=\int{s^\sigma\,d\tau},
\qquad \sigma\ne 1,
\end{equation}

\be
\n{-1}
z=\ln s,     \qquad    \eta=\int{\frac{d\tau}{s}},   \qquad \sigma=-1.
\ee

\noindent Then, in these new variables, the Eq. (\ref{s..}) becomes a linear
homogeneous differential equation with constant coefficients

\begin{equation}
\label{z}
\frac{d^2z}{d\eta^2}+\frac{dz}{d\eta}+\frac{\sigma+1}{4}z=0,
\qquad   \sigma\ne -1,
\end{equation}

\noindent equivalent to a damped harmonic oscillator equation, and

\be
\n{-2}
\frac{d^2z}{d\eta^2}+\frac{dz}{d\eta}+\frac{1}{4}=0,  \qquad  \sigma=-1.
\ee

\no On the other hand, expressing the Eq. (\ref{kgp}) for the \kf in terms of
the independent variable $\eta$, we get

\be
\n{ec}
\frac{d\phi}{\phi d\eta}=-\frac{3da}{\sigma ad\eta}+\frac{1}{2}.
\ee

\no Once $z(\eta)$ is known from Eq. (\ref{z}), one can compute $s(\tau)$ from
Eq. (\ref{zeta}), the scale factor $a(\tau)$ from (\ref{stau}), and the \kf
$\phi(\tau)$ from equation (\ref{kgp}). Following this procedure and inserting
the general solutions of Eqs. (\ref{z}) and (\ref{ec}) into the Einstein
equation (\ref{00}), we find the scale factor, the \kf and the relationship
amongst the integration constants

\begin{equation}
\label{ag1}
a=\left[\sqrt{-B}\,\mbox{e}^{-\frac{\eta}{2}}
\sinh{\left(\frac{\sqrt{-\sigma}}{2}\,\eta+
\eta_0\right)}\right]^{-\sigma/3(\sigma+1)}, \qquad \sigma<-1,
\end{equation}

\begin{equation}
\label{ag2}
a=a_0\mbox{e}^{-\eta/12+V_0\mbox{e}^{-\eta}/81c^2\phi_0^2},
\qquad \sigma=-1,
\end{equation}

\begin{equation}
\label{ag3}
a=\left[\sqrt{B}\,\mbox{e}^{-\frac{\eta}{2}}
\cosh{\left(\frac{\sqrt{-\sigma}}{2}\,\eta+
\eta_0\right)}\right]^{-\sigma/3(\sigma+1)}, \qquad -1<\sigma<0,
\end{equation}

\begin{equation}
\label{ag4}
a=\left[\sqrt{B}\,\mbox{e}^{-\frac{\eta}{2}}
\sin{\left(\frac{\sqrt{\sigma}}{2}\,\eta+
\eta_0\right)}\right]^{-\sigma/3(\sigma+1)}, \qquad 0<\sigma,
\end{equation}

\no the \kf is given by

\begin{equation}
\label{5.11}
\phi=\phi_0\, a^{-3/\sigma}\,\mbox{e}^{\eta/2},
\end{equation}

\noindent where

\be
\n{B}
B=\frac{4(\sigma+1)V_0}{27c^2\phi_0^2},
\ee

\no $\eta_0$ and $\phi_0$ are arbitrary integration constant.

For $\sigma<-1$, the solution expands from a singularity as $t^{1/3}$ and ends
as $t^{-\sigma/3}$. When $\sigma>-3$ the scale factor displays a power-law
inflationary scenario. For $-1<\sigma<0$, the universe expands from a
singularity as $t^{1/3}$ and its final behaviour is given by $t^{1/3}$. For
$\sigma>0$ the solution represents a contracting universe which begins at a
finite time, reaches a minimum where it bounces, exhibiting a final
superaccelerated expansion $a(t)\propto (t_0-t)^{-\sigma/3}$. The universe has
a finite time span and bounces when the \kf satisfies the condition
$\dot\phi^2=-1/m$.

As this model displays an accelerated expanding stage at late times it may be
an interesting alternative to describe the epoch where dark energy dominates.

%============================================================================
\subsection{The explicit solution}
%============================================================================

For $\sigma=-4$ or $mV_0=4/3$, we can solve the Eq. (\ref{s..}) by means of
the substitution

\begin{equation}
\label{24}
s^{-4}=\frac{1}{2}\frac{v^{-4}}{\int{v^{-4}\,d\tau}},
\end{equation}

\no so that Eq. (\ref{s..}) reads as

\begin{equation}
\label{25}
\ddot{v}=0.
\end{equation}

\noindent Inserting its solution into (\ref{24}), using (\ref{stau}) and
integrating (\ref{ec}), we get the general solution for the evolution and the
\kf satisfying the Eq. (\ref{00}):

\begin{equation}
\label{26}
a(\tau)=\left(\frac{3}{2}\right)^{1/3}
\left[\frac{9c^2V_0}{16\phi_0^2}t^4-ct\right]^{1/3},
\end{equation}

\begin{equation}
\label{27}
\phi(\tau)=\frac{2}{3t^{1/2}}
\left[-\frac{\phi_0^2}{c}+\frac{9V_0}{16}t^3\right]^{1/2},
\end{equation}

\noindent where we have chosen $c<0$ to set the singularity at $t=0$.
The scale factor (\ref{26}) exhibits a transition from $a\propto t^{1/3}$ to
an accelerated expansion where $a\propto t^{4/3}$. Curiously, it coincides
with the solutions found in \cite{bbarrow} where it was considered a FRW
spacetime filled with a scalar field driven by an exponential potential. In
the next section we will investigate this relation between the scalar field
and the \ke field. The \kf diverges at the singularity as $\phi\propto
t^{-1/2}$ and behaves as $\phi\propto t$ for large time. It has a minimum and
a turning point where the kinetic energy vanishes.

Another set of solutions can be found when $c=0$. Here, the Eq. (\ref{hp})
reduces to $\dot H=-H^2/mV_0$ and we get the power-law expansion $a=t^{mV_0}$
and a linear \kf $\phi=\phi_0 t$.

%============================================================================
\subsection{The polynomial function
$F_\gamma(x)=(-x)^{\frac{\gamma}{2(\gamma-1)}}$}
%============================================================================

This polynomial function yields a constant baryotropic index $\gamma$ and the
power-law expansion $a=t^{2/3\gamma}$ \cite{al}. For $F_\gamma$ the general
solution of Eq. (\ref{kgp}) is

\be
\n{sf}
\phi=\left[b+4\phi_0 t^{(2-\gamma)/\gamma}\right]^{\gamma/(2-\gamma)},
\qquad   \gamma\ne 2,
\ee

\be
\n{2}
\phi=\phi_0 t^b,      \qquad  \gamma=2,
\ee

\no where $b$ and $\phi_0$ are integration constants. Consistency between
these solutions and the Eq. (\ref{00}) gives the following relation between
the integration constants,

\be
\n{consis}
V_0=\frac{4(1-\gamma)}{3\gamma^2}(4\phi_0)^{\gamma/(1-\gamma)},
\qquad   \gamma\ne 2
\ee

\be
\n{bb}
b=\pm \frac{1}{\sqrt{3V_0}},    \qquad   \gamma=2.
\ee

\no For large cosmological time the \kf behaves as $\phi\approx t$. More
details about the cosmological model generated by the set $F_\gamma$ can be
found in Ref. \cite{al}.

%==================================================================
\subsection{Constant potential case}
%==================================================================

We have seen that the generalized Chaplygin gas model was proposed as unified
dark matter. It is derived from a Lagrangian containing non-standard
kinetic-energy terms (i.e., non-quadratic) and can be considered as driven by
a constant potential. Below we show that, even in absence of any potential
energy term, a general class of models exists, including the ECG, connecting a
dust dominated era at early times with an accelerated expansion stage at late
times.

For a constant potential $V=V_0$, the \kf equation (\ref{ga.}) has the first
integral

\be
\n{p0}
\frac{\gamma}{\dot\phi}=\frac{c}{a^3H^2},
\ee

\no or, after using Eqs. (\ref{ga}) and (\ref{11}), it turns into

\be
\n{pi0}
a^3F_x\dot\phi=\frac{3c}{2V_0},
\ee

\no where $c$ is an arbitrary integration constant. Also, combining them with
the Friedmann equation (\ref{00}), the baryotropic index associated with this
kind of \ke can be written in a more convenient form

\be
\n{gam}
\gamma=\frac{1}{1+2V_0^2a^6FF_x/9c^2}.
\ee

\no From the last equation we see that for a large set of model, i.e., the
class of model generated by the set of functions $F$ satisfying the condition
$a^6FF_x\ll 1$ at early time, the universe is dust dominated in the
beginning. At intermediate times it behaves as it were filled with a perfect
fluid with equation of state $p\propto\rho$. Finally, the universe ends in an
accelerated expansion scenario. So, these alternative models play the same
role that the generalized Chaplygin gas i.e., interpolating between dark
matter at early time and dark energy at late time.

Now, we investigate a simple kinetic \ke model generated by a functon $F$
satisfying the more general condition $a^6FF_x\approx constant$ at early time.
This model is generated by the following function

\be
\n{1}
F=\frac{1}{(2\alpha-1)V_0}\left[2\alpha \alpha_0\sqrt{-x}-(-x)^\alpha\right],
\ee

\no where $\alpha$ and $\alpha_0$ are two real constants. The energy density
and pressure of the \kf are calculated from Eqs. (\ref{ro}) and (\ref{1})

\be
\n{1f}
\rho=(-x)^\alpha,   \qquad   p=-\frac{1}{(2\alpha-1)}
\left[2\alpha \alpha_0\sqrt{-x}-(-x)^\alpha\right],
\ee

\no the equation of state is

\be
\n{1a}
p=\frac{1}{2\alpha-1}\left[\rho-
\frac{2\alpha \alpha_0}{\rho^{-1/2\alpha}}\right],
\ee

\no and the sound speed becomes

\be
\n{cs}
c_s^2=\frac{1}{2\alpha-1}\left[1-
\frac{\alpha_0}{\rho^{(2\alpha-1)/2\alpha}}\right].
\ee

\no Solving the conservation equation (\ref{con}), we obtain the energy
density

\be
\n{rom}
\rho=\left[\alpha_0+\frac{c_0}{a^3}\right]^{2\alpha/(2\alpha-1)},
\ee

\no where $c_0$ is a redefinition of the integration constant $c$. Restricting
to positive sound speed, the source (\ref{rom}) and the ECG induce a similar
evolution of the scale factor. However, there is a significant change when
both constant $\alpha_0$, $c_0$ are positive and $\alpha>1/2$, because in this
case the baryotropic index becomes

\be
\n{gapl}
\gamma=\frac{2\alpha}{2\alpha-1}\left[1-
\frac{\alpha_0}{\rho^{(2\alpha-1)/2\alpha}}\right].
\ee

\no So, near the singularity, the energy density diverges and the baryotropic
index behaves as $\gamma\approx 2\alpha/(2\alpha-1)$, indicating that the
model begins to evolve with a power-law dominated phase, where the scale
factor is $a\propto t^{(2\alpha-1)/3\alpha}$. In this limit the sound speed
(\ref{cs}) behaves as $c_s^2\approx (2\alpha-1)^{-1}$. For large $\alpha$, the
model is initially dust dominated, with approximated vanishing sound speed,
approaching to the ECG generated by the source (\ref{ro1c}) in the limit of
large $r$. At late times the model ends in a de Sitter stage. Such "modified
Chaplygin gas" may be considered as an alternative model to the generalized
Chaplygin gas investigated in \cite{bento-02}. It allows the evolution of the
initial perturbations in the energy density into a nonlinear regime to form a
gravitational condensate of particles that could play the role of cold dark
matter. The cool dark matter condensates gravitationally into the regions
where the pressure $p\approx 0$ and the \kf is close to
$\phi_c=(2\alpha\alpha_0)^{1/(2\alpha-1)}$. In this case, the model yields an
energy density which scales like the sum of a non-relativistic dust component
at $\phi=\phi_c$ with equation of state $p = 0$ and a
cosmological-constant-like component $p=-\rho$.

%============================================================
\section{Linking scalar and $k$-essence cosmologies}
%============================================================

The observed acceleration of the present Universe has been investigated
assuming that the dark energy can be described by quintessence and more
recently by \ke. The last one involves an effective scalar field theory
generated by a Lagrangian with a non-canonical kinetic term. Particular cases
of \ke are generalized Chaplygin gas and tachyon dark energy models.
Quintessence and \ke frameworks are usually based in a homogeneous scalar
fields driven by an exponential potential in the case of quintessence or an
inverse square potential in the case of k-essence. Both encounter the
so--called coincidence problem, namely, why are the energy densities of dark
energy and dark matter of the same order today. Standard quintessence model
appear promising in this point, as it can solve this problem for flat FRW
universes provided the dark matter component is assumed to be dissipative
\cite{overcome}. The system is attracted to a stationary and stable solution
characterized by the constancy of both density parameters at late times. In
addition, a class of k-essence models has been claimed to solve the
coincidence problem by linking the onset of dark energy domination to the
epoch of matter domination \cite{ar2,ar1}. From these satisfactory results and
taking into account that both models involve evolving scalar fields, we
believe it is reasonable to explore whether quintessence and \ke frameworks
have same kind of similitudes. We will deal with this question by
investigating in which cases quintessence resembes k-essence.

Another interesting aspect to be considered is related with current
observations which would indicate that the universe is superaccelerated and it could be
filled with a non-standard fluid that violates the weak energy condition. In
this sense, it was recently proposed a kind of matter described by an
homogeneous scalar field with negative kinetic energy term. The fluid with
negative pressure obeys and equation of state of the form $p=(\gamma-1)\rho$,
where $\gamma$ is taken negative and the models are known as phantom or ghost
cosmologies \cite{cal}-\cite{ruth}. The phantom and quintessence cosmologies
can be investigated simultaneously by taking a scalar field with both signs of
the kinetic term driven by an exponential potential. The dynamical equations
of these cosmological models are

\begin{equation}
\label{00q}
3H^{2}=\frac{1}{2}q\dot\varphi^{2}+{\cal V}
\end{equation}

\begin{equation}
\label{kgq}
\ddot\varphi+3H\dot\varphi+\frac{1}{q}\frac{d{\cal V}}{d\varphi}=0,
\end{equation}

\begin{equation}
\label{pe}
{\cal V}(\varphi)={\cal V}_0\mbox{e}^{-qA\varphi},
\end{equation}

\noindent where $q$, $A$ are real numbers and ${\cal V}_0$ is a positive
constant. The exponential potential is interesting because it may be
considered just as a limit of a more complex potential \cite{easther}. For
negative values of $q$ the above equations describe a phantom cosmology
\cite{ruth}.

It will be demonstrated that the scale factor obtained from Eqs.
(\ref{00q})-(\ref{pe}) or from the \ke model generated by the linear function
$F=1+mx$, and driven by an inverse square potential (see Eqs. (\ref{00}),
(\ref{V}), (\ref{hp}) and (\ref{fg})) is the same. To this end, we sketch the
procedure followed in the Ref. \cite{2c}. From Eqs. (\ref{00q})-(\ref{pe}), we
find

\begin{equation}
\label{9}
\dot{H}=-\frac{1}{2}q\dot\varphi^{2},
\end{equation}

\no and the first integral of the Klein-Gordon equation (\ref{kgq})

\begin{equation}
\label{3.91}
\dot\varphi =AH+\frac{c_1}{a^{3}},
\end{equation}

\no where $c_1$ is an arbitrary integration constant. Now, inserting Eq.
(\ref{3.91}) into (\ref{9}), we obtain the second-order differential equation
for the scale factor

\begin{equation}
\label{14}
\frac{d^2S}{d\zeta^2}+
S^{\nu}\frac{dS}{d\zeta}+\frac{1}{4}S^{2\nu+1}=0, \qquad \nu=-6/qA^2,
\end{equation}

\no where we have used the new variables $S$ and $\zeta$, defined by

\begin{equation}
\label{st}
S=a^{-3/\nu},   \qquad      \zeta=c_1qAt.
\end{equation}

\no With the following identification of the parameters

\be
\n{id}
mV_0=\frac{2}{qA^2},      \qquad \frac{3c}{2}=\frac{c_1}{A},
\ee

\no the Eqs. (\ref{s..}) and (\ref{14}) coincide. Therefore, both models are
described by the same scale factor and they are geometrically equivalent.
Also, from Eqs. (\ref{00},\ref{11}), (\ref{fg}) and Eqs.
(\ref{00q},\ref{kgq}), we get a relationship between both potentials

\be
\n{cons}
3H^2+\dot H=V(t)={\cal V}(t),
\ee

\no showing they are the same function of the cosmological time $t$, so that

\be
\n{i}
\frac{V_0}{\phi^2}={\cal V}_0\mbox{e}^{-qA\varphi}.
\ee

\no After inserting the \kf (\ref{5.11}) into the last equation we find the
scalar field

\be
\n{ce}
\varphi=\frac{1}{qA}\ln\varphi_0+\frac{\eta}{qA}+A\ln a,
\ee

\no where $\varphi_0={\cal V}_0\phi^2_0/V_0$. The Eq. (\ref{i}) supplies the
link between the scalar field and the \ke field

\be
\n{ii}
\phi=\frac{\phi_0}{\varphi_0^{1/2}}\mbox {e}^{qA\varphi/2},
\ee

\no displaying that these models are dynamically no equivalent. Thus, the
homogeneous quintessence and phantom fields are different than k-essence
field.

Resuming, from the cinematical point of view, that is the background geometry,
we have obtained exact equivalences and it is impossible to differentiate
between quintessence, k-essence and phantom cosmologies because they share the
same scale factor. To distinguish between them, it appears necessary to focus
our attention on the scalar field.

%==================================================================
\section{Conclusions}
%==================================================================

We have investigated the set of \Ts whose effective sound speed is
proportional to the sound speed of the tachyon field. They can be grouped into
three types according whether they yield phantom expansion or accelerated
expansion with or without inflation. We have shown these behaviours by finding
exact power-law solutions for an inverse square potential and proved that the
$T_{1/2}$-field is compatible only with this potential. Each \T produces an
ECG and the set of all these gases can be divided into three kinds, one of
which contains the generalized Chaplygin gas and the others give rise to
"perfect fluids", leading to new evolutions. There exists basically
nonsingular bouncing solutions which interpolate between two superaccelerated
stages or singular ones with a finite time span, as well as, peculiar singular
solutions that begin with a finite scale factor. In this manner, both, the ETF
and the ECG may be considered fair candidates to implement phantom
cosmologies.

For an inverse square potential, we have found the first integral of the \kf
equation for any function $F$ and shown that the coupled Einstein-\kf
equations can be solved in some cases. In particular, the divergent \ke theory
generated by the $T_{1/2}$-field becomes an intrinsic component of all \ke
models. Therefore, for power-law expansions the linear \kf model driven by an
inverse square potential and the divergent model are isomorphic. We have
obtained the general solution of Einstein-\kf equations for a linear function
$F$. From the kinematical point of view this model and the quintessence scalar
field one driven by an exponential potential are the same. However, they are
dynamically nonequivalent, because the \kf and the scalar field are linked by
the Einstein equation, e.g., both potentials are the same function of the
cosmological time.

For a constant potential, we have studied a \ke field associated with a
perfect fluid whose equation of state contains a term proportional to the
energy density and the other has the form of an ECG. This model, essentially
different than the ECG model, smoothly interpolates between a power-law
dominated phase and a de Sitter phase. In this "modified Chaplygin gas"
scenario, it may be chosen the value of the \kf where initial perturbations
condensate gravitationally in cold dark matter.

%==================================================================

%==================================================================

\section{acknowledgments}

This work was supported by the University of Buenos Aires under Project X223
and the Consejo Nacional de Investigaciones Cient\'{\i}ficas y T\'ecnicas.
LPC thanks Alejandro S. Jakubi for useful discussions of this work.

%\end{acknowledgments}

%==================================================================


\begin{thebibliography}{99}
%==================================================================

\bibitem{li}
A. R. Liddle and D. H. Lyth, {\it Cosmological Inflation and Large Scale
Structure}, Cambridge University Press (2000).

\bibitem {be} C. L. Bennett \textit{et al.},
astro-ph/0302207
\bibitem{Netterfield} C. B. Netterfield \textit{et al.},
astrophysics.J. \textbf{571}, 604 (2002)
\bibitem{ha} N. W. Halverson,
\textit{et al.},Astrophysics. J. \textbf{568}, 38 (2002)


\bibitem{ar3}
C. Armendariz-Picon, T. Damour and V.Mukhanov,
Phys.Lett. B {\bf 458}, 209 (1999).

\bibitem{a} G. W. Gibbons, Phys. Lett. B 537, 1 (2002).
\bibitem{alex}
A. Feinstein,
Physical Review D {\bf 66}, 063511 (2002).
\bibitem{padma}
T. Padmanabhan,
Physical Review D {\bf 66}, 021301(R) (2002).
\bibitem{sami} M. Sami, D. Chingangham and T. Qureshi, hep-th/0205179
\bibitem{} A. Frolov, L. Kofman and A. Linde, hep-th/0204187.
\bibitem{b} S. Sugimoto and S. Terashima, hep-th/0205085.

\bibitem{sen1} A. Sen, hep-th/0203211; hep-th/0203265; hep-th/0204143.

\bibitem{acce} A. G. Riess et al., Astron. J. \textbf{116}, 1009 (1998),
\texttt{astro-ph/9805201};
S. Perlmutter et al., Astrophys. J.
\textbf{517}, 565 (1998), \texttt{astro-ph/9812133};
G. Efstathiou et al., Mon. Not. Roy. Ast. Soc. \textbf{330}, L29
(2002), \texttt{astro-ph/0109152}.

\bibitem{gib} G. W. Gibbons, hep-th/0204008;
T. Padmanabhan, hep-th/0204150;
G. Shiu and Ira Wasserman, hep-th/0205003;
T. Padmanabhan and T. Roy Choudhury, hep-th/0205055;
D. Choudhury, D.Ghoshal, D. P. Jatkar and S. Panda, hep-th/0204204;
K. Hashimoto, hep-th/0204203;
H. S. Kim, hep-th/0204191;
S. Sugimoto, S. Terashima, hep-th/0205085;
J. A. Minahan, hep-th/0205098;
L. Cornalba, M. S. Costa and C. Kounnas, hep-th/0204261;
H. B. Benaoum, hep-th/0205140;
Xin-zhou Li, Jian-gang Hao and Dao-jun Liu, hep-th/0204252;
Yun-Song Piao, Phys. Rev. D {\bf 66}, 121301 (2002);
Jian-gang Hao and Xin-zhou Li, Phys. Rev. {\bf 68}, 043501 (2003);
S. Nojiri and S. D. Odintsov, Phys. Lett. B {\bf 571}, 1 (2003);
L.Raul W. Abramo, Fabio Finelli, Phys.Lett. B {\bf 575}, 165 (2003).

\bibitem{brief}
L. P. Chimento, A. S. Jakubi and D. Pav\'on,
Physical Review D {\bf 67}, 087302 (2003).


\bibitem{bento-03}
M. C. Bento, O. Bertolani and A. A. Sen,
Phys. Rev. D {\bf 67}, 063003 (2003).

\bibitem{1}
L.M.G. Beca, P.P. Avelino, J.P.M. de Carvalho and C.J.A.P. Martins
Phys.Rev. D {\bf 67}, 101301 (2003).

\bibitem{2}
N. Bilic, R. J. Lindebaum, G. B. Tupper, R. D. Viollier, astro-ph/0307214

\bibitem{3}
T. Multamaki, M. Manera and E. Gaztanaga,
Phys.Rev. D {\bf 69}, 023004 (2004).

\bibitem{4}
P.P. Avelino, L.M.G. Beca, J. P. M. de Carvalho, C.J.A.P. Martins and E.J.
Copeland,
Phys.Rev. D {\bf 69}, 041301 (2004).

\bibitem{5}
A. Dev, D. Jain and D. D. Upadhyaya, astro-ph/0311056

\bibitem{cat}
D.\ Carturan and F.\ Finelli,
Phys. Rev. D {\bf 68}, 103501 (2003);
L.\ Amendola,  F.\ Finelli, C.\ Burigana, and D.\
Carturan, JCAP 0307, 005 (2003)

\bibitem{zal} H.B.\ Sandvik, M.\ Tegmark, M.\ Zaldarriaga and I.\ Waga,
astro-ph/0212114.

\bibitem{ar2} C. Armendariz-Picon, V. Mukhanov, and Paul J. Steinhardt,
Phys. Rev. Lett. {\bf 85}, 4438 (2000).

\bibitem{ar1}
C. Armendariz-Picon, V. Mukhanov, and Paul J. Steinhardt,
Phys. Rev. D {\bf 63}, 103510 (2001).

\bibitem{mal1}
M.~Malquarti, E.~J.~Copeland, A.~R.~Liddle and M.~Trodden,
astro-ph/0302279.

\bibitem{car}
S.~M. Carroll, M.~Hoffman and M.~Trodden,
Phys. Rev. D {\bf 68}, 023509 (2003).

\bibitem{al}
L. P. Chimento and A. Feinstein,
(astro-ph/0305007).

\bibitem{kam}
A. Kamenshchik, U. Moschella and V. Pasquier,
Phys. Lett. {\bf B511}, 265 (2001).

\bibitem{bi-02}
N. Bilic, G. B. Tupper and R. D. Viollier
Phys. Lett. B {\bf 535}, 17 (2002);
N. Bilic, R. J. Lindebaum, G. B. Tupper and R. D. Viollier
astro-ph/0310181.

\bibitem{bento-02}
M. C. Bento, O. Bertolani and A. A. Sen,
Phys. Rev. D {\bf 66}, 043507 (2002).

\bibitem{sen}
A. Sen,
{\it JHEP} {\bf 0204}, 48 (2002).

\bibitem{gibb}
G. W. Gibbons,
Phys. Lett. {\bf B537}, 1 (2002).

\bibitem{obs}
S.~Hannestad and E.~M{\"o}rstell, Phys. Rev. D {\bf 66}, 063508 (2002);
P.H. Frampton and T.~Takahashi, Phys. Lett. B {\bf 557} 135 (2003);
S.M.~Carroll, M.~Hoffman and M.~Trodden,
Phys.\ Rev.\ D {\bf 68}, 023509 (2003);
A.~Melchiorri, L.~Mersini, C.J.~Odman, and M.~Trodden,
Phys.\ Rev.\ D {\bf 68}, 043509 (2003); J.S. Alca\~niz,  astro-ph/0312424;
 J.A.S. Lima, J.V. Cunha, and J.S. Alca\~niz,
Phys. Rev. D {\bf 68}, 023510 (2003).

\bibitem{str}
L.~Mersini, M.~Bastero-Gil and P.~Kanti, Phys.
Rev. D {\bf 64},  043508 (2001);
 M.~Bastero-Gil, P.H.~Frampton, and L.~Mersini,
Phys. Rev. D {\bf 65}, 106002 (2002);
P.H.~Frampton, Phys. Lett. B {\bf 555}, 139 (2003).

 \bibitem{fate} R.R.~Caldwell, M.~ Kamionkowski, N.N.~Weinberg,
Phys. Rev. Lett. {\bf 91}, 071301 (2003);  A.  Yurov,  astro-ph/0305019;  P.F. Gonz\'alez-D\'{\i} az, Phys. Rev. D {\bf 68}, 021303 (2003); M. Sami and A. Toporensky,  gr-qc/0312009, J.G. Hao, X.Z. Li, 
astro-ph/0309746; V. Sahni and Y. Shtanov, J. Cosm. Astro. Phys. {\bf 0311}, 014 (2003).

\bibitem{mal2}
M.~Malquarti, E.~J.~Copeland and A.~R.~Liddle,
Phys. Rev. D {\bf 68}, 023512 (2003).

\bibitem{lmath}
L. P. Chimento,
J. Math. Phys. {\bf 38}, 2565 (1997).

\bibitem{bbarrow}
A. B. Burd and J. D. Barrow,
Nucl. Phys. B {\bf 308}, 929 (1988).

\bibitem{overcome}
L.P. Chimento, A.S. Jakubi and D. Pav\'on,
Phys. Rev. D {\bf 62}, 063508 (2000); {\it ibid.} {\bf 67}, 087302 (2003).

\bibitem{cal}
R.R.~Caldwell, Phys. Lett. B {\bf 545}, 23 (2002).

\bibitem{Mel:2002ux}
A.~Melchiorri, L.~Mersini, C.~J.~Odman and M.~Trodden,
astro-ph/0211522.

\bibitem{Car:2003st}
S.~M.~Carroll, M.~Hoffman and M.~Trodden,
Phys.\ Rev.\ D {\bf 68}, 023509 (2003)

\bibitem{easther}
R. Easther,
Class. Quantum Grav. {\bf 10} 2203 (1993).

\bibitem{ruth}
L. P. Chimento and R. Lazkoz,
Phys. Rev. Lett. {\bf 91} 211301 (2003).

\bibitem{2c}
L. P. Chimento,
Class. and Quantum Grav. {\bf 15}, 965 (1998).



\end{thebibliography}
\end{document}